\documentclass[10pt,journal, twocolumn]{IEEEtran}

\usepackage[utf8x]{inputenc}
\usepackage{cite}
\usepackage{graphicx}
\usepackage{amsmath}
\usepackage{algorithmic}
\usepackage{bm}
\usepackage{array}
\usepackage{url}
\usepackage[justification=centering]{caption}
\usepackage{subfigure}
\usepackage{fancyhdr}
\title{An End-to-End Block Autoencoder For Physical Layer Based On Neural Networks}
\author{Tianjie Mu, Xiaohui Chen, Li Chen, Huarui Yin, and Weidong Wang}

\begin{document}
	\maketitle
	\thispagestyle{fancy} 
	\lhead{} 
	\chead{} 
	\rhead{} 
	\lfoot{} 
	\cfoot{} 
	\rfoot{\thepage} 
	\renewcommand{\headrulewidth}{0pt} 
	\renewcommand{\footrulewidth}{0pt} 
	\pagestyle{fancy}
	\rfoot{\thepage}
	\begin{abstract}
		Deep Learning has been widely applied in the area of image and natural language processing. In this paper, we propose an end-to-end communication structure based on autoencoder where the transceiver can be optimized jointly. A neural network roles as a combination of channel encoder and modulator. In order to deal with input sequences parallelly, we introduce block scheme, which means that the autoencoder divides the input sequence into a series of blocks. Each block contains fixed number of bits for encoding and modulating operation. Through training, the proposed system is able to produce the modulated constellation diagram of each block. The simulation results show that our autoencoder performs better than other autoencoder-based systems under additive Gaussian white noise (AWGN) and fading channels. We also prove that the bit error rate (BER) of proposed system can achieve an acceptable range with increasing the number of symbols.
	\end{abstract}
	\section{introduction}
	In the past, conventional methods optimize the modules of communication system separately, such as encoder, modulator, to achieve the better transmission quality\cite{boche2004outage}\cite{el2004lattice}. Deep Learning has experienced fast development in the past decade and it also possesses great potential in wireless communication. There have been lots of model-driven applications based on Deep Learning\cite{wang2017deep}, such as massive MIMO\cite{huang2018deep} and OFDM\cite{felix2018ofdm}.
	\par
	One important application of Deep Learning is to view communication system as an end-to-end autoencoder, in which the modules can be optimized jointly. The result in\cite{o2017introduction} has shown that autoencoders can readily match the performance of nearoptimal existing baseline modulation and coding schemes by learning the system during training. The transmitter maps a one-hot vector to particular constellation symbols for transmission. The signals distorted by channel are used to reconstruct the original vector. The authors of \cite{raj2018backpropagating} have proved that an end-to-end structure need a differential channel model to optimize the transceiver. However, one-hot transmission scheme is limited because all information bits are only used to transmit one symbol, which decreases transmission efficiency seriously.
	\par
	Opposite to one-hot transmission scheme, block scheme is a transmission scheme which allows parallel inputs. It enables communication systems to transmit a stream of information bits instead of bits for one symbol\cite{o2018physical}. In\cite{zhu2019joint}, block scheme\cite{jones1994block}is introduced in autoencoder to deal with the transmission of batches of sequences. This structure supports arbitrary length of binary sequences as input, but its performance is not good enough for practical use.\par
	In this paper, we build up an end-to-end autoencoder with block transmission scheme. In order to improve its performance, we also introduce memory mechanism into the neural networks. Our contributions are following:
	\begin{itemize}
		\item We propose a novel autoencoder structure based on neural networks. It introduces block scheme to deal with sequences in the form of blocks and allows arbitrary input length, which improves transmission efficiency. With the memory mechanism of recurrent neural networks (RNN), the autoencoder explores potential relationships between blocks for modulating. Through optimizing the transmitter and receiver jointly, the constellation diagram can be learned automatically for particular modulation mode.
		\item We train and test the model under different channel models. The performance of the proposed model is better than other autoencoder-based communication systems under typical channels\cite{zhu2019joint}. At the same time, the simulation result shows that lower code rate leads to a lower bit error rate (BER).
	\end{itemize}
	\section{deep neural network structures}
	A deep feedforward network, which is also called multi-layer perceptron, is a typical deep learning model\cite{goodfellow2016deep}. Feedforward networks define a map $y = f(\boldsymbol{x;\theta})$, and use backpropagation\cite{hecht1992theory} to learn the value of $\theta$, obtaining the best nonlinear approximation of some function $f^{*}(\boldsymbol{x})$ we need. There is no feedback between the output and the model itself. When there exists connection, it is called recurrent neural network(RNN).
	\par
	Given a particular amount of training samples, we send them into the networks as batches. The output is used to calculate the loss and compute the gradient. The computed gradient is broadcast back through the neural networks and the parameter vector $\boldsymbol{\theta}$ is update according to the gradient.
	\par
	There are several typical kinds of layers of neural networks.
	\begin{itemize}
		\item Fully-connected layer. Its neural units between two adjacent layers are fully-connected. Each neural unit has an activation function to introduce nonlinearity into the network such as $ReLU$ and $sigmoid$. Therefore the fully-connected layer has a strong ability to approximate $f^{*}(\boldsymbol{x})$.
		\item  Convolutional neural networks consist a series of filters called kernel. The kernels generates receptive field and extract features of input like images. Convolutional networks have been applied in some novel communication structures. In\cite{o2017introduction}, CNNs accomplish classification tasks for different modulation schemes.
		\item A long short-term memory (LSTM) network is an artificial RNN architecture. It introduces memory mechanism and extracts relationships between time steps. LSTM can learn to bridge minimal time lags in excess of 1000 discrete-time steps by enforcing constant error flow through constant error carousels within special units\cite{hochreiter1997long}. The architecture of LSTM we adopt in our system is shown in Fig.1. Cells are connected recurrently to each other, replacing the usual hidden units of ordinary recurrent neural networks. An input feature is computed with a regular artificial neuron unit. Its value can be accumulated into the state if the sigmoidal input gate allows it. The state unit has a linear self-loop whose weight is controlled by the forget gate. The output of the cell can be shut off by the output gate. All the gating units have a sigmoidal nonlinearity, while the input unit can have any squashing nonlinearity\cite{goodfellow2016deep}.
	\end{itemize}

	\begin{figure}[!ht]
		\centering
		\includegraphics[width=0.6\linewidth]{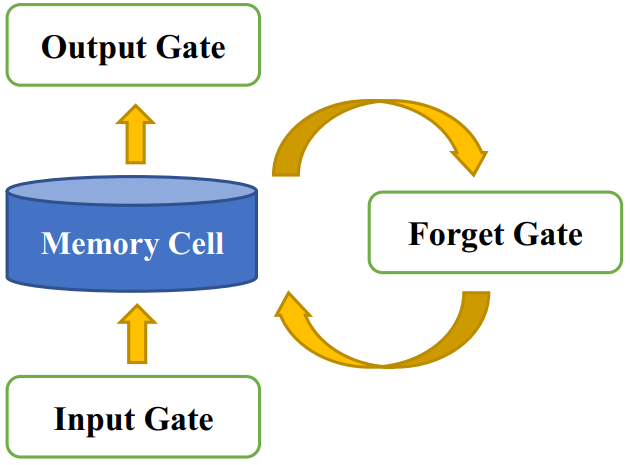}
		\caption{A common structure of LSTM network.}
		\label{fig:fig1}
	\end{figure}
	\section{system model}
	We build up an end-to-end communication system using neural networks feeding with block data, which enables us to complete joint optimization of transceiver.
	\subsection{Network Structure}
	\par
	The structure of block autoencoder is shown in Fig.2. It consists several parts as following.
	\begin{itemize}
		\item The input is a stream of bits. To solve the problem of block transmission, we set the number of blocks to $M$, and each block has $S$ bits to be modulated, so the total length of input bits is $S\times M$.
		\item In the first layer, we adopt a convolutional neural network to compress input bits into $M$ blocks. The output is sent to several LSTM layers to produce the modulated $M$ complex symbols. We combine the time distributed layer with LSTM layer in order to introduce some linear relationship between symbols. To satisfy power constraint, we normalize the output symbols at the end of the transmitter. The detailed parameters of our autoencoder are shown in table \uppercase\expandafter{\romannumeral1}.
		\item Since we add the operation of encoding into the network through adjusting output dimension of time-distributed layers, the number of complex symbols should be $M'$ instead of $M$, which is dependent on the code rate we set.
	\end{itemize}	
	\begin{figure}[!ht]
		\centering
		\includegraphics[width=0.8\linewidth]{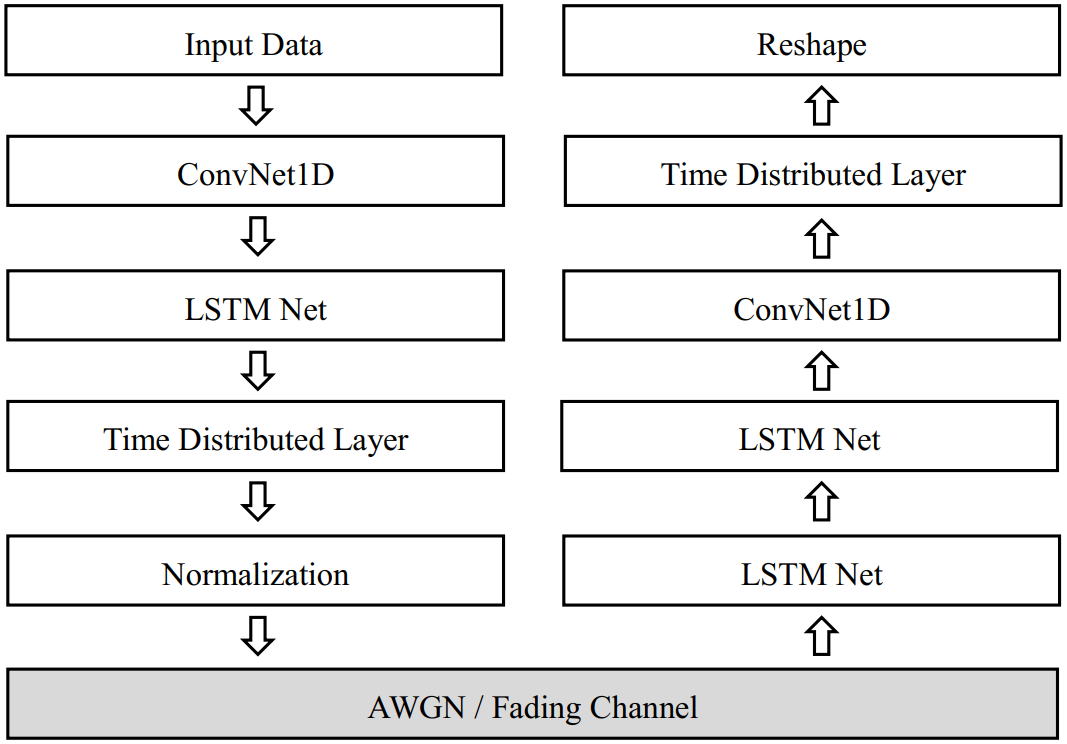}
		\caption{The structure of proposed autoencoder based on LSTMs and CNNs.}
		\label{fig:fig2}
	\end{figure}
	\begin{table}
		\renewcommand{\arraystretch}{1.3}
		\caption{Detailed Parameters of Block Autoencoder}
		\label{table_example}
		\centering
		\begin{tabular}{c||c}
			\hline
			\bfseries Layer & \bfseries Parameters\\
			\hline\hline
			Input & $S \times M \ \{0,1\}$\\
			\hline
			Conv1D\_1 & stride=S, kernel=S, filters=128\\
			\hline
			LSTM\_1 & units=400\\
			\hline
			Time Distributed & $(M', 2)$\\
			\hline
			LSTM\_2 & units=128\\
			\hline
			LSTM\_3 & units=64\\
			\hline
			Conv1D\_2 & stride=1(default), kernel=S, filters=64\\
			\hline
			Time Distributed & $(M, S)$\\
			\hline
			Reshape & $S \times M$\\
			\hline
		\end{tabular}
	\end{table}
	\par
	Following the encoding and modulating operation, the coded sequence $\bm{z}$ is transmitted over the communication channel by I and Q components of digital signal. In our model, the communication channel is non-trainable, which can be represented as $h(\bm{z})$.
	\par
	The distorted signal $\bm{z}' \in{\mathbf{C^{M'}}}$ is demodulated and decoded by the receiver. These layers reconstruct the input sequence. Each trainable layer of proposed autoencoder is followed by a batch normalization layer so that the training process will converge more quickly.
	\subsection{Channel model}
	\begin{itemize}
	\item First we consider AWGN channel models. AWGN channel is used to train and test our autoencoder. We add zero-mean complex Gaussian noise to the transmitted symbol $\bm{z}$. The variance of noise is calculated by given $E_{b}/N_{0}$ and block size $S$.
	\par
	 \item In wireless communication, frequency selective fading is a radio propagation anomaly caused by partial cancellation of a radio signal by itself. The signal arrives at the receiver by several different paths. There exists inter-symbol interference (ISI) that influences the signal to be received. For generalization, we also do experiments under frequency selective fading channels. Traditional methods add protective interval to avoid or decrease ISI. However, our autoencoder is an end-to-end system, so we simply increase the number of symbols instead of introducing extra artificial symbols into the end of transmitter. We train and test the models under two multi-path channels. The channel models we use are shown in Fig.3. Channel A has two fading paths and the zero-delayed path is strong. Different from channel A, channel B has three fading paths, including a weak zero-delayed one.
   \end{itemize}
	\begin{figure}[!ht]
	\centering
	\includegraphics[width=0.8\linewidth]{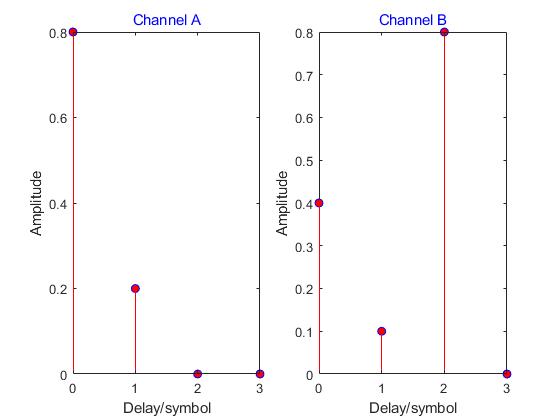}
	\caption{These two fading channel models are chosen to do experiments with.}
	\label{fig:fig3}
	\end{figure}
%
	\section{experiments}
	In order to obtain the performance of proposed autoencoder, we train and test the model in different scenarios. Bit error rate (BER) is a measure of the number of bit errors that occur in a given number of bit transmissions under all scenarios. For generalization, we simply select AWGN channel model. In fact, under the scenario of wireless communication, the channel would be more complex because signals arrive at the receiver through different paths which leads to ISI between symbols.
	\subsection{Settings}
	For simulation, we set the block size to 6 and block number to 400. So the autoencoder acts like a joint coding and modulating 64-QAM system. We compare the learned autoencoder with conventional coding and modulating method. The data sets are generated by random distributed $\{0,1\}$. The number of samples is 40000 for training and 10000 for testing. We set batch size to 64 and use Adam optimizer with learning rate 0.001. We need to train the autoencoder under an SNR-fixed channel. Through several experiments, we find the best training $E_{b}/N_{0}$ is 12dB.
	\begin{figure}[ht]
		\centering
		\includegraphics[width=0.8\linewidth]{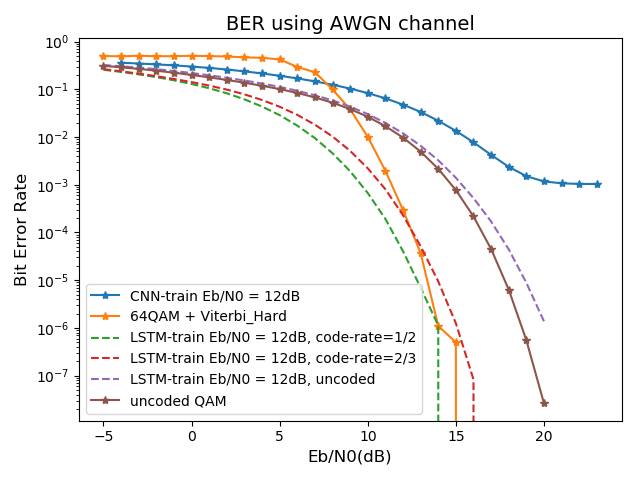}
		\caption{The comparison between proposed autoencoder and autoencoder based on CNNs\cite{zhu2019joint}}
		\label{fig:fig4}
	\end{figure}	
	\subsection{AWGN Channel}
	The performance of the autoencoder under AWGN channel is shown in Fig.4. We also implement the autoencoder in  \cite{zhu2019joint} for comparison. We add redundant information to resist the influence of channel through increasing the number of symbols. The way that we adjust the code rate is to set different dimension to the time-distributed layer and the convolutional layer in the decoder. When code rate is set to 1, which means the sequence is uncoded, our autoencoder performs very closely to conventional MMSE decoding method. Clearly as shown in Fig.5, our block autoencoder gives better performance than autoencoder in \cite{zhu2019joint}. When we decrease the code rate to $2/3$, which means we add redundant information to the encoded sequence, the autoencoder's performance is improved rationally. When code rate is set to $1/2$, we compare it with Viterbi hard decoding method in 64QAM. We can find that our autoencoder performs far beyond Viterbi hard decoding method in low SNR situation. It requires lower power to reach the same BER as Viterbi's method.

	\begin{figure}[ht]
		\centering
		\includegraphics[width=0.8\linewidth]{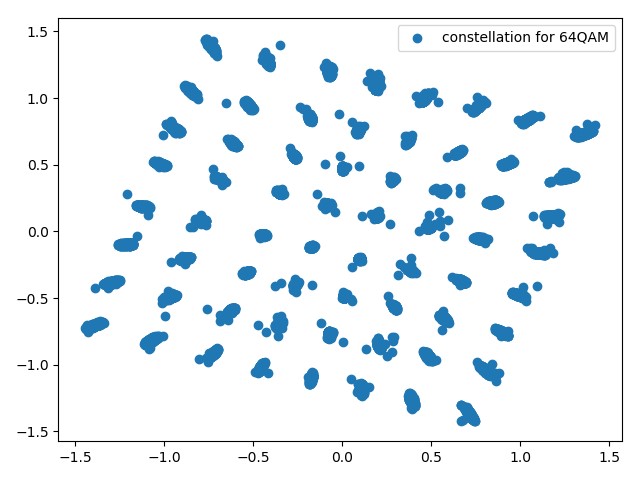}
		\caption{The learned constellation diagram for 64 QAM.}
		\label{fig:fig5}
	\end{figure}

	\par We draw the constellation diagram of the trained autoencoder in Fig.5. We can see that the symbols plotted in complex plane are distributed in 64 clusters. In actual deployment, it is easy to transfer symbols through inphase and quadrature component according to the constellation diagram.
	
	\subsection{Fading Channel}
	The performance under two chosen channels is shown in Fig.6. We set the code rate to 1/2 and training $E_{b}/N_{0}$ to 20dB. Our autoencoder performs well in the noise ranging from -5dB to 10dB but faces an error floor when $E_{b}/N_{0}$ is more than 15dB. Compared with channel A, channel B's BER is higher because it contains a weaker zero-delay path is weaker.
	\begin{figure}[ht]
		\centering
		\includegraphics[width=0.8\linewidth]{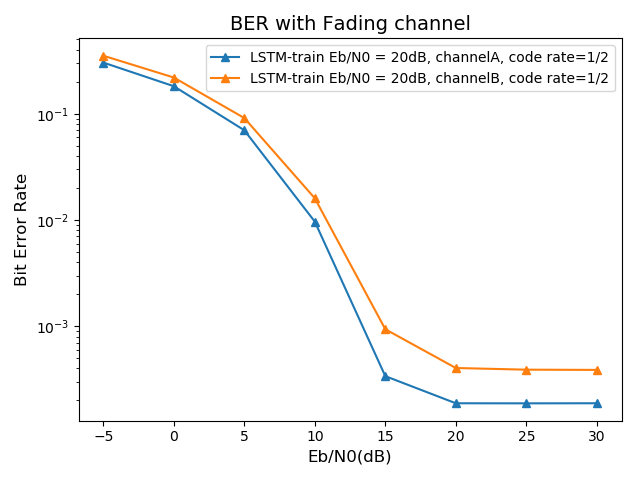}
		\caption{The performance of block autoencoder under fading channels.}
		\label{fig:fig6}
	\end{figure}
	\par
	To improve the autoencoder's performance, we continue to decrease the code rate. As shown in Fig.7, its BER decreases when we amplify the number of symbols under the same channel B when we set training BER to 12dB. However, this will reduce the transmission efficiency so that the system is hard to be deployed on hardware. So trade-off strategy is important.
	\begin{figure}[ht]
		\centering
		\includegraphics[width=0.8\linewidth]{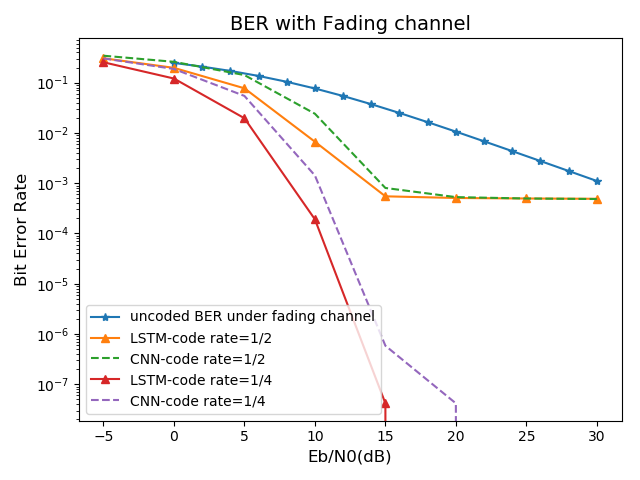}
		\caption{The performance of block autoencoder with different code rates.}
		\label{fig:fig7}
	\end{figure}
	\section{conclusion}
	In this paper, we propose a new communication structure combined with LSTMs and CNNs. The autoencoder performs better than other autoencoder-based communication systems under AWGN and multi-path fading channels. A regular constellation diagram can be learned with the limit of average power, which is easier to be deployed on hardware platform. Considering the wireless transmission scenario, the autoencoder needs extra symbols to resist the channel fading. The simulation result shows that the BER of proposed autoencoder can be decreased to an acceptable range through reducing code rate. We show that we can decrease the code rate to ensure a satisfying BER. Due to the property of CNNs and LSTMs, the autoencoder has no limit on the length of input sequence. Furthermore, we prove that the training and testing process do not need a particular channel model.
	\par
	We may further discover other applications based on the block autoencoder in the following aspects.
	\begin{itemize}
		\item Our autoencoder is a kind of SISO system. The spectrum efficiency of SISO system is much lower than MIMO\cite{el2004lattice}. MIMO systems can enhance throughput without more bandwidth or transmit power expenditure. MIMO has become an essential element of wireless communication standards including IEEE 802.11n (Wi-Fi), IEEE 802.11ac (Wi-Fi), HSPA+ (3G), WiMAX (4G), and Long Term Evolution (4G LTE). Therefore, it is necessary for us to extend our system to a MIMO autoencoder.
		\item We mention that we can increase the number of symbols to reach to an ideal BER range. For proposed autoencoder, however, the code rate should be low to achieve the acceptable performance, which means we need to add more redundant information. So it is important to design a better structure based on block autoencoder, which shows more robustness to fading channels.
	\end{itemize}

	\bibliographystyle{IEEEtran}
	\bibliography{blockAE0615}

\begin{thebibliography}{10}
\providecommand{\url}[1]{#1}
\csname url@samestyle\endcsname
\providecommand{\newblock}{\relax}
\providecommand{\bibinfo}[2]{#2}
\providecommand{\BIBentrySTDinterwordspacing}{\spaceskip=0pt\relax}
\providecommand{\BIBentryALTinterwordstretchfactor}{4}
\providecommand{\BIBentryALTinterwordspacing}{\spaceskip=\fontdimen2\font plus
\BIBentryALTinterwordstretchfactor\fontdimen3\font minus
  \fontdimen4\font\relax}
\providecommand{\BIBforeignlanguage}[2]{{%
\expandafter\ifx\csname l@#1\endcsname\relax
\typeout{** WARNING: IEEEtran.bst: No hyphenation pattern has been}%
\typeout{** loaded for the language `#1'. Using the pattern for}%
\typeout{** the default language instead.}%
\else
\language=\csname l@#1\endcsname
\fi
#2}}
\providecommand{\BIBdecl}{\relax}
\BIBdecl

\bibitem{boche2004outage}
H.~Boche and E.~A. Jorswieck, ``Outage probability of multiple antenna systems:
  Optimal transmission and impact of correlation,'' in \emph{International
  Zurich Seminar on Communications, 2004}.\hskip 1em plus 0.5em minus
  0.4em\relax IEEE, 2004, pp. 116--119.

\bibitem{el2004lattice}
H.~El~Gamal, G.~Caire, and M.~O. Damen, ``Lattice coding and decoding achieve
  the optimal diversity-multiplexing tradeoff of mimo channels,'' \emph{IEEE
  Transactions on Information Theory}, vol.~50, no.~6, pp. 968--985, 2004.

\bibitem{wang2017deep}
T.~Wang, C.-K. Wen, H.~Wang, F.~Gao, T.~Jiang, and S.~Jin, ``Deep learning for
  wireless physical layer: Opportunities and challenges,'' \emph{China
  Communications}, vol.~14, no.~11, pp. 92--111, 2017.

\bibitem{huang2018deep}
H.~Huang, J.~Yang, H.~Huang, Y.~Song, and G.~Gui, ``Deep learning for
  super-resolution channel estimation and doa estimation based massive mimo
  system,'' \emph{IEEE Transactions on Vehicular Technology}, vol.~67, no.~9,
  pp. 8549--8560, 2018.

\bibitem{felix2018ofdm}
A.~Felix, S.~Cammerer, S.~D{\"o}rner, J.~Hoydis, and S.~Ten~Brink,
  ``Ofdm-autoencoder for end-to-end learning of communications systems,'' in
  \emph{2018 IEEE 19th International Workshop on Signal Processing Advances in
  Wireless Communications (SPAWC)}.\hskip 1em plus 0.5em minus 0.4em\relax
  IEEE, 2018, pp. 1--5.

\bibitem{o2017introduction}
T.~O’Shea and J.~Hoydis, ``An introduction to deep learning for the physical
  layer,'' \emph{IEEE Transactions on Cognitive Communications and Networking},
  vol.~3, no.~4, pp. 563--575, 2017.

\bibitem{raj2018backpropagating}
V.~Raj and S.~Kalyani, ``Backpropagating through the air: Deep learning at
  physical layer without channel models,'' \emph{IEEE Communications Letters},
  vol.~22, no.~11, pp. 2278--2281, 2018.

\bibitem{o2018physical}
T.~J. O'Shea, T.~Roy, N.~West, and B.~C. Hilburn, ``Physical layer
  communications system design over-the-air using adversarial networks,'' in
  \emph{2018 26th European Signal Processing Conference (EUSIPCO)}.\hskip 1em
  plus 0.5em minus 0.4em\relax IEEE, 2018, pp. 529--532.

\bibitem{zhu2019joint}
B.~Zhu, J.~Wang, L.~He, and J.~Song, ``Joint transceiver optimization for
  wireless communication phy using neural network,'' \emph{IEEE Journal on
  Selected Areas in Communications}, 2019.

\bibitem{jones1994block}
A.~E. Jones, T.~A. Wilkinson, and S.~Barton, ``Block coding scheme for
  reduction of peak to mean envelope power ratio of multicarrier transmission
  schemes,'' \emph{Electronics letters}, vol.~30, no.~25, pp. 2098--2099, 1994.

\bibitem{goodfellow2016deep}
I.~Goodfellow, Y.~Bengio, and A.~Courville, \emph{Deep learning}.\hskip 1em
  plus 0.5em minus 0.4em\relax MIT press, 2016.

\bibitem{hecht1992theory}
R.~Hecht-Nielsen, ``Theory of the backpropagation neural network,'' in
  \emph{Neural networks for perception}.\hskip 1em plus 0.5em minus 0.4em\relax
  Elsevier, 1992, pp. 65--93.

\bibitem{hochreiter1997long}
S.~Hochreiter and J.~Schmidhuber, ``Long short-term memory,'' \emph{Neural
  computation}, vol.~9, no.~8, pp. 1735--1780, 1997.

\end{thebibliography}

\end{document}